# Dipole-matter interactions governed by the asymmetry of Maxwell's equations

Yuhan Zhong[1,2], Chan Wang[1,2,3], Chenxu Bian[1,2], Xuhuinan Chen[1,2], Jialin Chen[1,2,4], Xingjian Zhu[5], Hao Hu[6], Tony Low[7,8], Hongsheng Chen[1,2,3,9,*], Baile Zhang[10,11], and Xiao Lin[1,2,*]

[1]*Interdisciplinary Center for Quantum Information, State Key Laboratory of Extreme Photonics and Instrumentation, ZJU-Hangzhou Global Scientific and Technological Innovation Center, College of Information Science and Electronic Engineering, Zhejiang University, Hangzhou 310027, China.*
[2]*International Joint Innovation Center, Key Lab. of Advanced Micro/Nano Electronic Devices & Smart Systems of Zhejiang, the Electromagnetics Academy at Zhejiang University, Zhejiang University, Haining 314400, China.*
[3]*Jinhua Institute of Zhejiang University, Zhejiang University, Jinhua 321099, China.*
[4]*Department of Electrical and Computer Engineering, Technion-Israel Institute of Technology, Haifa 32000, Israel.*
[5]*School of Physics, Zhejiang University, Hangzhou 310027, China.*
[6]*School of Electrical and Electronic Engineering, Nanyang Technological University, Singapore 639798, Singapore.*
[7]*Department of Electrical and Computer Engineering, University of Minnesota, Minneapolis, Minnesota 55455, USA.*
[8]*Department of Physics, University of Minnesota, Minneapolis, Minnesota 55455, USA.*
[9]*Shaoxing Institute of Zhejiang University, Zhejiang University, Shaoxing 312000, China.*
[10]*Division of Physics and Applied Physics, School of Physical and Mathematical Sciences, Nanyang Technological University, Singapore 637371, Singapore.*
[11]*Centre for Disruptive Photonic Technologies, Nanyang Technological University, Singapore 637371, Singapore.*
[*]*Corresponding authors. Email: xiaolinzju@zju.edu.cn (X. Lin); hansomchen@zju.edu.cn (H. Chen)*

**Directionally molding the near-field and far-field radiation lies at the heart of nanophotonics and is crucial for applications such as on-chip information processing and chiral quantum networks. The most fundamental model for radiating structures is a dipolar source located inside a homogeneous matter. However, the influence of matter on the directionality of dipolar radiation is oftentimes overlooked, especially for the near-field radiation. We show that the dipole-matter interaction is intrinsically asymmetric and does not fulfill the duality principle, originating from the inherent asymmetry of Maxwell's equations, i.e., electric charge and current are ubiquitous but their magnetic counterparts are non-existent to elusive. Moreover, we find that the asymmetric dipole-matter interaction could offer an enticing route to reshape the directionality of not only the near-field radiation but also the far-field radiation. As an example, both the near-field and far-field radiation directionality of Huygens dipole**

**(located close to a dielectric-metal interface) would be reversed, if the dipolar position is changed from the dielectric region to the metal region.**

Maxwell's equations are the foundation of classical electromagnetics and photonics [1]. Recalling the Ampère's and Faraday's laws, namely

$$\nabla \times \bar{E}(\bar{r}) = i\omega \bar{B}(\bar{r}) \tag{1}$$

$$\nabla \times \bar{H}(\bar{r}) = -i\omega \bar{D}(\bar{r}) + \bar{J}(\bar{r}) \tag{2}$$

one makes the observation that Maxwell's equations are asymmetric and do not fulfill the duality principle, since they account for the contribution of an electric current $\bar{J}(\bar{r})$ but not the magnetic current [1]. For electromagnetic radiation, the electric current in Maxwell's equations represents the sole radiation source, among which the most fundamental ones are the electric and magnetic dipoles. The electric and magnetic dipoles are important notions for electromagnetic wave theory, and the manipulation of dipolar radiation can be traced back to the earliest days of electromagnetism, to the Hertzian dipole [2] and the Purcell effect [3-5]. In the 19$^{th}$ century, Hertz conceived the Hertzian dipole and used it to prove the existence of electromagnetic waves conclusively [2]. In the 1940s, Purcell discovered the Purcell effect and showed that the matter in which the dipolar source is embedded has a significant influence on the intensity of dipolar radiation [3]. Despite the extensive researches in dipolar radiation, the manipulation of its directionality, especially in the near-field, remains a topic of continual interest. Interest in this subject intensifies in recent years, as the ability to mold the flow of light in the extreme nanometer scale holds great promise for many enticing applications, such as the development of novel integrated light sources [6-8], optical communications [9-11], on-chip information processing [12, 13], and chiral quantum networks [14, 15]. However, the influence of dipole-matter interaction on the directionality of near-field radiation has been largely un-explored [16-21]. In this work, we show that the asymmetry in Maxwell equations facilitates a unique form of dipole-matter interactions, which do not fulfill the duality principle but allow for reshaping the directionality of both the near- and far-filed radiations.

The manipulation of the near-field radiation pattern of dipolar sources has so far been exclusively focused on two following approaches. One approach relies on the constituent design of dipolar sources [22-25]. For example, the circularly-polarized dipole with a spinning dipole-moment has a highly asymmetric near-field radiation pattern, whose directionality is governed by the spin-momentum locking [26-28]. By contrast, the Huygens dipole, composed of two orthogonal electric and magnetic dipole-moments, can have its near-field directionality determined by the time-averaged Poynting vector [22, 29-31]. Alternative approach relies on the geometrical design of neighboring out-couplers [32-35] (e.g., asymmetric waveguides [32] and irregular scatters [33-35]). As an example, the directionality of near-field radiation could be flipped by changing the polarization [36, 37] or the group velocity [38, 39] of surface waves supported by the out-coupler.

In this work, we propose a new paradigm to reshape the directionality of both near-field and far-field radiation, by exploiting the dipolar-matter interaction, under the scenario that both the dipolar constituents and the out-coupler are fixed. We find that the optical response of matter, including its permittivity and permeability, can strongly influence the spatial-frequency spectrum of waves carried by the dipolar source. As a result, we show that the radiation directionality of both transverse-magnetic (TM, or *p*-polarized) surface waves and propagating waves excited by Huygens dipoles would be flipped, if this dipole is close to a dielectric-metal interface but its position is changed from the dielectric region to the metal region. Moreover, we find that this exotic capability to reshape the radiation directionality is inherently limited by the choice of the dipolar source and out-coupler, due to the asymmetry of dipole-matter interactions. On the one hand, the directional near-field radiation would be insensitive to the dipole-matter interaction, if the source is constructed solely by electric dipole-moments (e.g., circularly-polarized electric dipoles) or solely by magnetic dipole-moments (e.g., circularly-polarized magnetic dipoles), since the variation of matter has the same influence on the spatial-frequency spectrum of all electric or magnetic dipole-moments. On the other hand, the dipole-matter interaction even becomes incapable to flip the radiation directionality for arbitrary dipolar sources if the out-coupler supports transverse-electric (TE, or *s*-polarized) surface waves. This incapability intrinsically comes from the fact that there is only the electric current in Maxwell's equations, while the magnetic current is missing.

We begin with Maxwell's equations in an isotropic matter in Fig. 1. From the constitution relation [1], we have $\bar{D}(\bar{r}) = \varepsilon_0 \varepsilon_r \bar{E}(\bar{r})$ and $\bar{B}(\bar{r}) = \mu_0 \mu_r \bar{H}(\bar{r})$, where $\varepsilon_0$ and $\mu_0$ are the permittivity and permeability in free space, respectively; $\varepsilon_r$ and $\mu_r$ represent the relative permittivity and permeability of matter, respectively. Correspondingly, equations (1-2) can be re-organized as

$$\nabla \times \bar{E}(\bar{r}) = i\omega\mu_0\mu_r \bar{H}(\bar{r}) \tag{3}$$

$$\nabla \times \bar{H}(\bar{r}) = -i\omega\varepsilon_0\varepsilon_r \bar{E}(\bar{r}) + \bar{J}(\bar{r}) \tag{4}$$

For dipolar sources, they are closely related to the electric current $\bar{J}(\bar{r})$. For example, the electric dipole with a moment of $\bar{p}_e = \hat{\alpha}p_e^\alpha$ ($\alpha = x, y, z$) can be treated as an electric current-carrying element with an infinitesimal length $l$; correspondingly, we have $-i\omega p_e^\alpha = I_e^\alpha l$ and $\bar{J}(\bar{r}) = \hat{\alpha}I_e^\alpha l \delta(\bar{r})$, with $I_e^\alpha$ being the electric current. According to the Ampère model [40], the magnetic dipole with a moment of $\bar{m}_e = \hat{\beta}m_e^\beta$ ($\beta = x, y, z$) can be effectively modelled by an electric current loop with an infinitesimally small radius $a$ in Fig. 1; then we have $m_e^\beta = I_e(\pi a^2)$ and $|\bar{J}(\bar{r})| = I_e \delta(r-a)\delta(\beta)$, where $I_e$ is the magnitude of electric current.

With the knowledge of $\bar{J}(\bar{r})$ in Maxwell's equations, the spatial-frequency spectrum of both TM and TE waves carried by arbitrary dipolar sources can be analytically derived. Below we use $E_z(x,y,z) = \iint_{-\infty}^{\infty} E_z(k_x, k_y, z) e^{ik_x x} e^{ik_y y} dk_x dk_y$ to describe TM waves, where we have $k_x^2 + k_y^2 + k_z^2 = \varepsilon_r\mu_r k_0^2$ in the matter, $k_0 = \omega/c$, $\omega$ is the angular frequency, and $c$ is the speed of light in free space; see the discussion of TE waves later. After some calculations [supporting section S1], we have

$$E_{\text{z}}(k_{\text{x}}, k_{\text{y}}, z) = \frac{i}{8\pi^2 \varepsilon_0} \left( \mp k_{\text{x}} \frac{p_{\text{e}}^{\text{x}}}{\varepsilon_{\text{r}}} \mp k_{\text{y}} \frac{p_{\text{e}}^{\text{y}}}{\varepsilon_{\text{r}}} + \frac{k_{\text{x}}^2 + k_{\text{y}}^2}{k_{\text{z}}} \frac{p_{\text{e}}^{\text{z}}}{\varepsilon_{\text{r}}} \right) e^{ik_{\text{z}}|z|} + \frac{i\omega\mu_0}{8\pi^2} \left( \frac{k_{\text{y}}}{k_{\text{z}}} \mu_{\text{r}} m_{\text{e}}^{\text{x}} - \frac{k_{\text{x}}}{k_{\text{z}}} \mu_{\text{r}} m_{\text{e}}^{\text{y}} \right) e^{ik_{\text{z}}|z|} \quad (5)$$

The dipole-matter interaction is now mathematically manifested in equation (5). That is, the modulation of the optical response of matter, including the change of $\varepsilon_{\text{r}}$ or $\mu_{\text{r}}$, would have a direct impact on the magnitude of field in equation (5) and thus the spatial-frequency spectrum in the *k*-space of dipolar sources as shown in Fig. 2. Moreover, equation (5) indicates that the modulation on matter's $\varepsilon_{\text{r}}$ is only capable of changing the radiation from the electric dipole-moment $p_{\text{e}}^{\alpha}$, while matter's $\mu_{\text{r}}$ is only related to the magnetic dipole-moment $m_{\text{e}}^{\beta}$. These intrinsic connections between the optical responses of matter and the TM field carried by the dipolar source indicate a promising route to engineering the radiation directionality of TM waves.

By following this thought, Fig. 2 shows the possibility of modulating the TM spatial-frequency spectrum of dipolar sources by exploiting the dipole-matter interaction. Here, the matter is set to be a non-magnetic material with $\mu_{\text{r}} = 1$, and the source is the Huygens dipole, which has its electric and magnetic dipole-moments satisfying Kerker's condition of $\frac{|\overline{m}_{\text{e}}|}{|\overline{p}_{\text{e}}|} = c$ [41-43]. To be specific, the Huygens dipole is composed of $p_{\text{e}}^{\text{z}} = p_0$ and $m_{\text{e}}^{\text{y}}/c = p_0$, where $p_0 = 1$ Coulomb·meter is a constant. When $\varepsilon_{\text{r}} > 0$, such as a dielectric with $\varepsilon_{\text{r}} = 1$ in Fig. 2a-b, we always have $\left|\frac{E_{\text{z}}(-|k_{\text{x}}|)}{E_{\text{z}}(+|k_{\text{x}}|)}\right| > 1$ (here $k_{\text{x}}$ is a real number) for the carried waves, indicating that the spatial-frequency spectrum of the Huygens dipole is highly asymmetric with respect to $k_{\text{x}}$. Conversely, if $\varepsilon_{\text{r}} < 0$, such as a metal with $\varepsilon_{\text{r}} = -2$ in Fig. 2c-d, we have $\left|\frac{E_{\text{z}}(-|k_{\text{x}}|)}{E_{\text{z}}(+|k_{\text{x}}|)}\right| < 1$. From Fig. 2b&d, we can further argue that the optical response of matter plays a key role in determining the asymmetry of the spatial-frequency spectrum, and this spectral asymmetry could be reversed by only changing the matter from $\varepsilon_{\text{r}} > 0$ to $\varepsilon_{\text{r}} < 0$.

To manifest the tunability of the spectral asymmetry in Fig. 2, the two matters used in Fig. 2 are chosen to construct a dielectric-metal interface in Fig. 3. This out-coupler supports TM surface waves with an in-plane wavevector of $k_{\text{sw,TM}} = 1.4k_0$ and its distance to the Huygens dipole is $d_0 = 0.4\lambda_0$, where $\lambda_0 = 2\pi/k_0$ is the wavelength of light in free space. When the Huygens dipole is inside the dielectric region (namely inside a matter with $\varepsilon_r > 0$), the excited TM surface waves and propagating waves mainly propagate to the $-x$ direction in Fig. 3a-b. By contrast, if the Huygens dipole is changed to be inside the metal region (namely inside a matter with $\varepsilon_r < 0$), most excited TM surface waves and propagating waves would flow to the $+x$ direction in Fig. 3c-d. Therefore, Fig. 3 shows that it is possible to reshape the directionality of both near-field and far-field radiation purely by changing the dipole-matter interaction.

Motivated by the exotic results in Figs. 2-3, it is natural to ask about the extent that the dipole-matter could flip the radiation directionality, especially for the near-field radiation. By further analyzing equation (5), we find that the capability to flip the radiation directionality requires both electric and magnetic dipole-moments in the source (e.g., the Huygens dipole). In other words, if the dipolar source is constructed solely by electric or magnetic dipole-moments (e.g., circularly-polarized electric dipole), the change in the optical response of matter would not significantly alter the near-field directionality; see Fig. S1.

We now discuss the influence of dipole-matter interactions on the radiation directionality of TE waves. Similarly, $H_z(x,y,z) = \iint_{-\infty}^{\infty} H_z(k_x, k_y, z) e^{ik_x x} e^{ik_y y} dk_x dk_y$ is adopted to describe TE waves carried by the dipolar source, where

$$H_z(k_x, k_y, z) = \frac{i\omega}{8\pi^2}\left(-\frac{k_y}{k_z}p_e^x + \frac{k_x}{k_z}p_e^y\right)e^{ik_z|z|} + \frac{i}{8\pi^2}\left(\mp k_x m_e^x \mp k_y m_e^y + \frac{k_x^2 + k_y^2}{k_z}m_e^z\right)e^{ik_z|z|} \quad (6)$$

Counterintuitively, the optical response of matter, including both $\varepsilon_r$ and $\mu_r$, is *not* directly related to any dipole-moment in equation (6). Therefore, the radiation of TE waves should be insensitive to the dipole-matter interaction, completely different from that of TM waves. That is, the directionality of excited TE surface waves and propagating waves is irrelevant to the optical response of matter, as exemplified in Fig. S2.

Upon close inspection, we find that the independence of radiation directionality of TE waves on the dipole-matter interaction revealed in equation (6) or Fig. S2 originates from the asymmetry of Maxwell's equations. That is, there is only the electric current $\bar{J}(\bar{r})$ existing in the asymmetric Maxwell equations, while the magnetic current $\bar{M}(\bar{r})$ is missing; see equations (3-4). Notably, the magnetic current is a useful concept, frequently used by applying the equivalence principle [1], although it may not exist in reality. If we add a hypothetical magnetic current to Faraday's law in Fig. S3, the originally-asymmetric Maxwell equations in equations (3-4) are changed to

$$\nabla \times \bar{E}(\bar{r}) = i\omega\mu_0\mu_r\bar{H}(\bar{r}) - \bar{M}(\bar{r}) \tag{7}$$

$$\nabla \times \bar{H}(\bar{r}) = -i\omega\varepsilon_0\varepsilon_r\bar{E}(\bar{r}) + \bar{J}(\bar{r}) \tag{8}$$

The modified Maxwell equations in equations (7-8) [1] become symmetric and fully satisfy the duality principle, by making the following replacements: $\bar{E}(\bar{r}) \to \bar{H}(\bar{r})$, $\bar{H}(\bar{r}) \to -\bar{E}(\bar{r})$, $\varepsilon_0 \to \mu_0$, $\varepsilon_r \to \mu_r$, $\mu_0 \to \varepsilon_0$, $\mu_r \to \varepsilon_r$, $\bar{J}(\bar{r}) \to \bar{M}(\bar{r})$, and $\bar{M}(\bar{r}) \to -\bar{J}(\bar{r})$. With the simultaneous existence of electric and magnetic currents in the symmetric Maxwell equations, both magnetic and electric dipoles can now be directly treated as magnetic and electric current-carrying elements with an infinitesimal length $l$, respectively. Then according to the Gilbert model [40], we have $-i\omega m_m^\beta = I_m^\beta l$ and $\bar{M}(\bar{r}) = \hat{\beta} I_m^\beta l \delta(\bar{r})$ with $I_m^\beta$ being the magnetic current; similarly, $-i\omega p_e^\alpha = I_e^\alpha l$ and $\bar{J}(\bar{r}) = \hat{\alpha} I_e^\alpha l \delta(\bar{r})$ with $I_e^\alpha$ being the electric current.

With the knowledge of both $\bar{J}(\bar{r})$ and $\bar{M}(\bar{r})$ in the symmetric Maxwell equations, the spatial-frequency spectrum of both TM and TE waves carried by arbitrary dipolar sources can be re-derived, and the corresponding field expressions in equations (5-6) should be changed to

$$E_z(k_x, k_y, z) = \frac{i}{8\pi^2\varepsilon_0}\left(\mp k_x \frac{p_e^x}{\varepsilon_r} \mp k_y \frac{p_e^y}{\varepsilon_r} + \frac{k_x^2+k_y^2}{k_z}\frac{p_e^z}{\varepsilon_r}\right)e^{ik_z|z|} + \frac{i\omega}{8\pi^2}\left(\frac{k_y}{k_z}m_m^x - \frac{k_x}{k_z}m_m^y\right)e^{ik_z|z|} \tag{9}$$

$$H_z(k_x, k_y, z) = \frac{i\omega}{8\pi^2}\left(-\frac{k_y}{k_z}p_e^x + \frac{k_x}{k_z}p_e^y\right)e^{ik_z|z|} + \frac{i}{8\pi^2\mu_0}\left(\mp k_x \frac{m_m^x}{\mu_r} \mp k_y \frac{m_m^y}{\mu_r} + \frac{k_x^2+k_y^2}{k_z}\frac{m_m^z}{\mu_r}\right)e^{ik_z|z|} \tag{10}$$

Remarkably, equations (9-10) which mathematically describe the dipole-matter interaction also fulfill the duality principle, by making the following replacements: $E_z \to H_z$, $H_z \to -E_z$, $\varepsilon_0 \to \mu_0$, $\varepsilon_r \to \mu_r$, $\mu_0 \to \varepsilon_0$, $\mu_r \to \varepsilon_r$, $p_e^{x,y,z} \to m_m^{x,y,z}$, and $m_m^{x,y,z} \to -p_e^{x,y,z}$. Moreover, the matter's permittivity (permeability) is directly related to the electric (magnetic) dipole-moments for the excitation of TM (TE) waves. Under this fictious scenario, controlling dipole-matter interactions should now be capable to reshape the radiation directionality of both TM and TE waves; see Figs. S4-S5.

Moreover, equations (9-10) would reduce to equations (5-6), if $m_\mathrm{m}^\beta = \mu_0\mu_\mathrm{r} m_\mathrm{e}^\beta$. Actually, the two magnetic dipoles of $m_\mathrm{m}^\beta$ and $m_\mathrm{e}^\beta$ used in this work have different système international (SI) units and differ by a factor of permeability $\mu_0\mu_\mathrm{r}$. To be specific, $m_\mathrm{e}^\beta$ defined by a current loop according to the Ampère model in Fig. 1 directly produces the $\bar{H}$ field (see equation (4)), while $m_\mathrm{m}^\beta$ defined by a pair of opposite magnetic charges according to the Gilbert model in Fig. S2 produces the $\bar{B}$ field (see equation (7)). As a result, the induced total fields in equations (5-6) and in equations (9-10) have distinct dependences on the variation of matter's permeability. This causes the difference for the asymmetric and symmetric dipole-matter interactions.

In conclusion, we have revealed that the dipole-matter interaction is intrinsically asymmetric and does not fulfill the duality principle, originating from the absence of magnetic currents in Maxwell's equations. We have found that the asymmetric dipole-matter interaction can be exploited to reshape the directionality of both the near-field and far-field radiation of dipolar sources. As an example, the radiation directionality of Huygens dipole would be flipped, if it is deposited close to a dielectric-metal interface but its position is changed from the dielectric to the metal. The underlying reason is that regulating the optical response of matter is capable to flexibly tailor the spatial-frequency spectrum of waves carried by the dipolar sources. Upon close inspection, we have found that this enticing capability, as governed by the asymmetry of Maxwell's equations, is dependent on the judicious choice of the dipolar source and the out-coupler. Our findings thus may provide an extra degree of freedom to reshaping the light flow both in the near and far fields by applying the asymmetric dipole-matter interaction.

**Acknowledgements**
X.L. acknowledges the support partly from the National Natural Science Fund for Excellent Young Scientists Fund Program (Overseas) of China, the National Natural Science Foundation of China (NSFC) under Grant No. 62175212,

Zhejiang Provincial Natural Science Fund Key Project under Grant No. LZ23F050003, and the Fundamental Research Funds for the Central Universities (2021FZZX001-19). H.C acknowledges the support partly from the Key Research and Development Program of the Ministry of Science and Technology under Grants No. 2022YFA1404704, 2022YFA1404902, and 2022YFA1405200, and the National Natural Science Foundation of China (NNSFC) under Grants No.11961141010 and No. 61975176. J.C. acknowledges the support from the Chinese Scholarship Council (CSC No. 202206320287). B.Z. acknowledges the support from National Research Foundation Singapore Competitive Research Program no. NRF-CRP23-2019-0007, Singapore Ministry of Education Academic Research Fund Tier 3 under grant no. MOE2016-T3-1-006 and Tier 2 under grant no. MOE2019-T2-2-085.

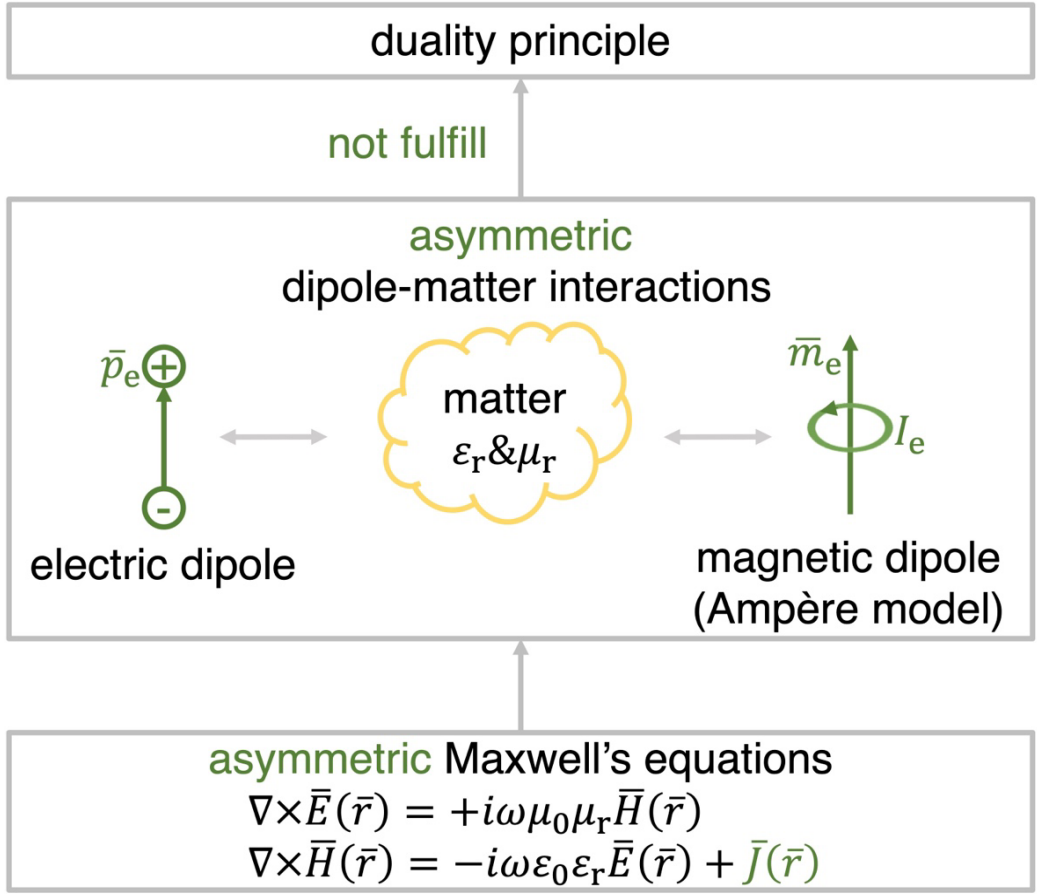

**FIG. 1. Asymmetric dipole-matter interactions.** Since the magnetic current is missing in Maxwell's equations, both electric dipole and magnetic dipole are solely related to the electric current density $\bar{J}(\bar{r})$. The dipole-matter interaction intrinsically does not fulfill the duality principle, as governed by the asymmetric Maxwell equations.

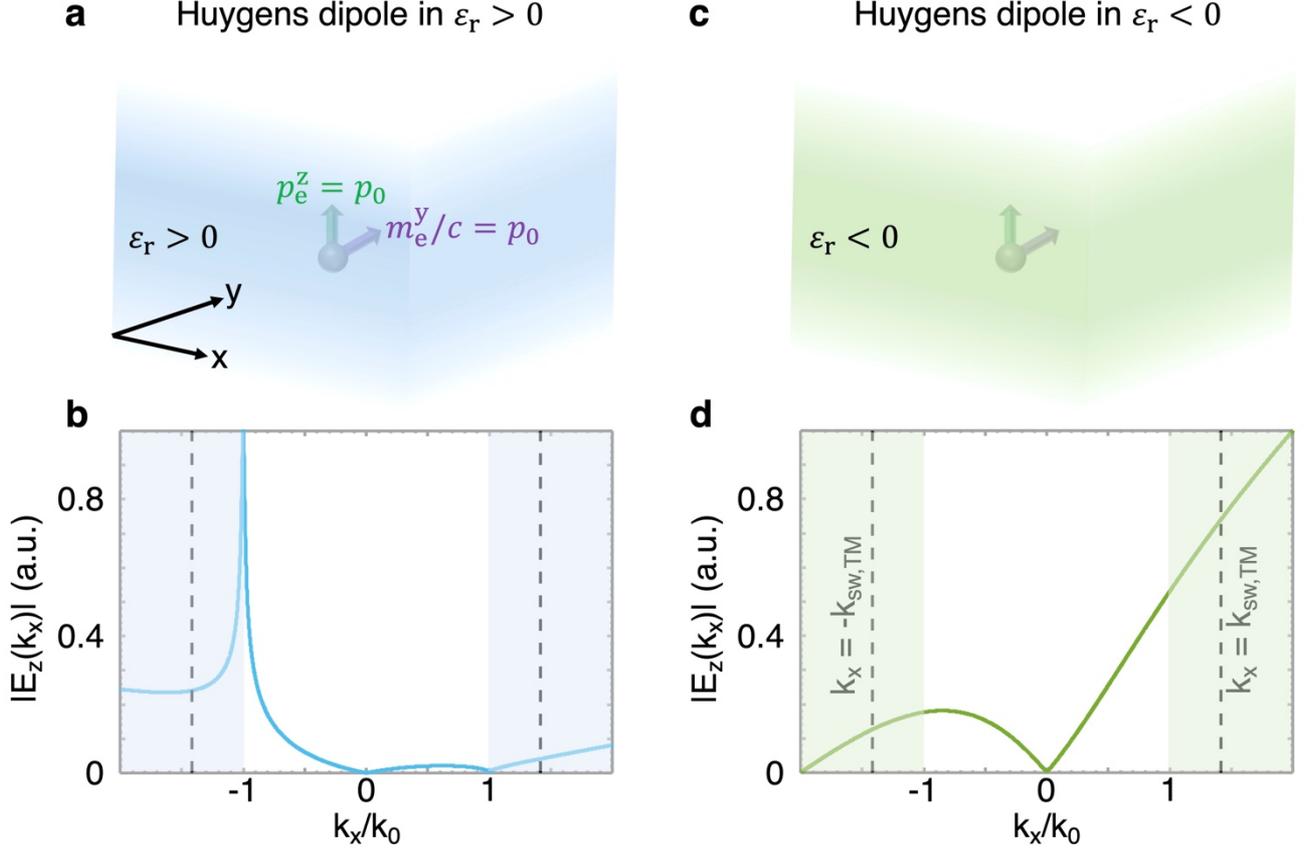

**FIG. 2. Matter-dependent spatial-frequency spectrum of TM waves carried by the dipolar source.** Here the dipolar source is the Huygens dipole, composed of $p_e^z = p_0$ and $m_e^y/c = p_0$, and it is located inside a matter with a relative permittivity of $\varepsilon_r$. (**a, b**) Matter with $\varepsilon_r = 1$. (**c, d**) Matter with $\varepsilon_r = -2$. Due to the dipole-matter interaction, we have $\left|\frac{E_z(-|k_x|)}{E_z(+|k_x|)}\right| > 1$ if $\varepsilon_r > 0$, no matter $\left|\frac{k_x}{k_0}\right| > 1$ (corresponding to evanescent waves in free space) or $\left|\frac{k_x}{k_0}\right| < 1$ (corresponding to propagating waves in free space). Conversely, $\left|\frac{E_z(-|k_x|)}{E_z(+|k_x|)}\right| < 1$ if $\varepsilon_r < 0$. In each plot of the spatial-frequency spectrum, we set $k_y = 0$ for conceptual brevity here and below.

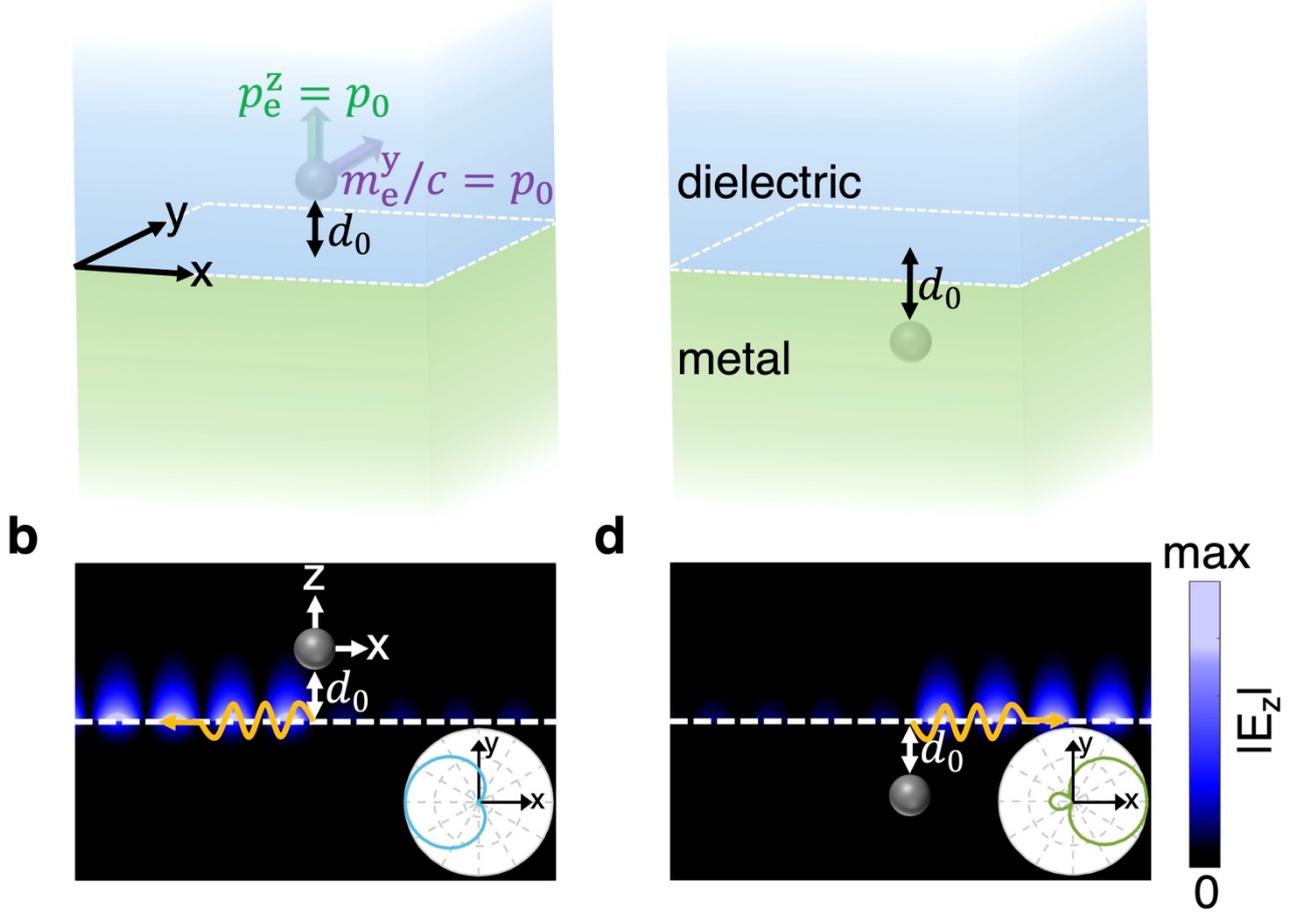

**FIG. 3. Dipole-matter interaction to reshape the directionality of both near-field and far-field radiation.** The dipolar source is the Huygens dipole, and the out-coupler is a dielectric-metal interface, which supports the propagation of TM surface waves with an in-plane wavevector of $k_{sw,TM} = 1.4k_0$. The relative permittivities of dielectric and metal are $\varepsilon_{r,diel} = 1$ and $\varepsilon_{r,metal} = -2$, respectively. The source-interface distance is $d_0 = 0.4\lambda_0$, where $\lambda_0$ is the wavelength of light in free space. (**a, b**) Huygens dipole inside a matter with $\varepsilon_r > 0$, namely the dielectric region. (**c, d**) Huygens dipole inside a matter with $\varepsilon_r < 0$, namely the metal region. The insets in (b, d) show the far-field radiation patterns of Huygens dipole in the plane of $z = 2d_0$. Due to the dipole-matter interaction, both the excited surface waves and propagating waves mainly propagate along the $-x$ ($+x$) direction if the dipole is inside the matter with $\varepsilon_r > 0$ ($\varepsilon_r < 0$).